\documentclass[aps,prb,twocolumn,showpacs,floatfix]{revtex4}
\usepackage{latexsym}
\usepackage{graphicx}
\newcommand{\figwidth}{3.275 in}
\usepackage{subfigure}    
\usepackage{epsfig}
\begin{document}
\title{Simulation of melting of two dimensional Lennard-Jones solids}
\author{Keola Wierschem$^1$ and Efstratios Manousakis$^2$}
\affiliation{
$^1$School of Physical and Mathematical Sciences, Nanyang Technological University, 
21 Nanyang Link, Singapore 637371\\
$^2$Department of Physics, Florida State University,
Tallahassee, FL 32306-4350, USA\\
and Department of  Physics, University of Athens,
Panepistimioupolis, Zografos, 157 84 Athens, Greece
}
\date{\today}
\begin{abstract}
We study the nature of melting of a two dimensional (2D) Lennard-Jones
solid using large scale Monte Carlo simulation. We use systems of 
up to 102,400 particles to capture the decay of the correlation functions 
associated with translational order (TO) as well as the
bond-orientational (BO) order. 
We  study the role of dislocations and disclinations
and their  distribution functions. We computed the temperature
dependence of the second moment of the TO order parameter
($\Psi_G$)  as well as of the order parameter $\Psi_6$ associated with
BO order. Applying finite-size
scaling of these second moments we determined the anomalous dimension
critical exponents $\eta(T)$ and $\eta_6(T)$ associated with power-law 
decay of the $\Psi_G$ and $\Psi_6$ correlation functions. We
also computed the temperature dependent distribution of the order
parameters $\Psi_G$ and $\Psi_6$ on the complex plane which support a
two stage melting with a hexatic phase as an intermediate phase.
From the correlation functions of $\Psi_G$ and $\Psi_6$ we extracted the 
corresponding temperature dependent
correlation lengths $\xi(T)$ and $\xi_6(T)$. The analysis of our
results leads to a consistent picture strongly supporting  a
two stage melting scenario as predicted by the Kosterlitz, Thouless, 
Halperin, Nelson, and Young (KTHNY) theory where melting
occurs via  two continuous  phase transitions,  first from
solid  to  a hexatic  fluid at temperature $T_m$, and  then  from the hexatic
fluid to  an isotropic fluid at a critical temperature $T_i$.  We find that
$\xi(T)$ and $\xi_6(T)$  have a distinctly different temperature
dependence each diverging at different temperature and that their
finite size scaling properties are consistent with the KTHNY theory.
We also used the temperature dependence of $\eta$ and $\eta_6$ and their
theoretical bounds to provide estimates for the critical temperatures
$T_m$ and $T_i$, which can also be estimated using the Binder ratio.
Our results are within error bars the same as
those extracted from the divergence of the correlation lengths.

\end{abstract}
\pacs{64.60.De,67.70.D-,61.72.Bb}
\maketitle

\section{Introduction}

The most widely considered theory of 2D melting is the  so-called KTHNY  theory of Kosterlitz  and Thouless \cite{KT},  
Halperin and Nelson \cite{HN,NH}, and  Young \cite{Young}, which  
predicts that melting  in two
dimensions  occurs via  two continuous  phase transitions,  first from
solid  to hexatic  fluid, and  then  from hexatic  fluid to  isotropic
fluid. This theory begins from the fact that true translational order
cannot exist at any non-zero temperature in 2D because of the infrared
divergence caused by the zero point motion of long-wave-length density
fluctuations. According to the KTHNY theory, another form of true
long-range order exists below some non-zero temperature $T_m$ where
only the directions of the nearest-neighbor bonds order. This
long-range bond order disappears above $T_m$ because of dislocation
unbinding which leads to an intermediate phase, the hexatic phase,
characterized by topological order, where while dislocations are
unbound, disclinations with opposite topological charge 
remain bound. These disclinations become unbound at a higher
temperature $T_i$ where the system becomes an isotropic disordered fluid.
 
Simulation of melting in classical two-dimensional (2D) systems has
been tackled by means of a variety of computational studies\cite{Krauth} for
several decades without reaching a definite conclusion regarding its
nature. In particular for hard disks in 2D, a large number of computer
simulation studies have been applied to understand 2D melting, since
this is the toy model on which the Metropolis Monte Carlo method
itself was first introduced\cite{Metropolis} and soon afterward,  the 2D melting of
hard disks was studied\cite{Disks}. 
One of the reasons for the difficulty to reach an
unequivocal conclusion is that in 2D a conventional solid
with true translational order cannot exist, and, instead the
correlations decay very slowly over long distance. This requires large
size systems where the relaxation time
scales become very long for these types of phenomena. In particular
for hard disk systems, when using a local updating algorithm or even molecular
dynamics, particles remain stuck in their local ``cage'' for large computational
time scales, precisely because of the hard disk constraint.

 One might think that Monte Carlo simulation of soft-core potentials,
 such as  the Lennard-Jones  system in  2D,  might not be plagued by
 the same  level of computational severity as the hard-disk
systems, because of the softening of the hard-core constraint.
As a matter of fact there are a number of studies of the Lennard-Jones
solid\cite{Chester} by computer simulation where also a general
consensus about the nature of melting has not been established. Some
studies  have favored a  first-order transition  from solid  to liquid
\cite{Toxvaerd,Abraham,Bakker,Chester84},  as predicted by  the grain 
boundary melting suggestion\cite{Chui},  while other studies 
\cite{Frenkel,Udink,Somer,Somer98}  have leaned  toward the KTHNY
theory. The most thorough of these studies, however, are at least one
decade old and because of the fact that the computational resource
constraints of today are significantly better, a more thorough study  should be possible.
  
In the present paper, we study the nature of melting of a two 
dimensional (2D) Lennard-Jones solid using large scale Monte Carlo 
simulation.  We use systems of  up to 102,400 particles to capture the 
decay of the correlation functions  associated with translational as 
well as the bond-orientational order.  We find that to carry out 
thorough investigations beyond these sizes, calculations using the 
Metropolis local update become impractical using today's high 
performance computing because  of the long relaxation time scales. 
Further  technical details of our simulation are described in the next section, 
and the remainder of the paper is organized as follows. 
In Sec.~\ref{defects} we discuss the role of defects in the KTHNY theory 
of melting and present the results of a geometric defect analysis. 
In Sec.~\ref{orders} we show the temperature dependence of both order 
parameters, $\Psi_{G}$ and $\Psi_{6}$, as well as their second moments, 
$\Psi_{G}^{2}$ and $\Psi_{6}^{2}$. 
The system-size dependence  of $\Psi_{G}^{2}$ and $\Psi_{6}^{2}$ can be 
used to determine the critical exponents $\eta$ and $\eta_{6}$, as shown 
in Sec.~\ref{exponents}. In the same section, the KTHNY values of the 
critical exponents at melting, $\eta(T_{m})$ and $\eta_{6}(T_{i})$, are 
used to estimate the transition temperatures $T_{m}$ and $T_{i}$. 
Next, in Sec.~\ref{correlations}, we present our results on the correlation
function associated with bond orientational order above $T_i$ and
determine the temperature dependent correlation length $\xi_6(T)$. 
In the same section, we demonstrate finite-size scaling of $\Psi_{6}^{2}$. 
A similar presentation is given in Sec.~\ref{distributions} for the 
pair distribution function and the correlation length of translational order, 
$\xi(T)$. In addition, we present our findings for the scaling behavior 
of the second moment of $\Psi_{G}^{2}$ in this same section. 
Sec.~\ref{binder} presents an analysis of the melting 
transition using Binder's cumulant ratio~\cite{Binder81} for each order 
parameter, and also includes a discussion of finite-size scaling theory 
in the presence of multiple correlation lengths. 
Finally, in Sec.~\ref{conclusions}, we give a brief summary of our 
main findings and conclusions.

\section{Simulation details}
\label{methods}

 In the Lennard-Jones  potential,  for two particles separated by a 
distance $r$,
\begin{equation}
V(r)=4 \epsilon \left( \left(\frac{\sigma}{r}\right)^{12} - \left(\frac{\sigma}{r}\right)^{6} \right),
\end{equation}
an attractive  inverse sixth power  tail is combined with  a repulsive
inverse  twelfth power  hard core, such that there are only  
two parameters: $\epsilon$, the
potential  well   depth,  and  $\sigma$,  the  hard-core   diameter.  
However, our results can be inferred for any particular value of these
parameters (for a specific real system), since in 
 our calculations, distance  is  measured  in  units  of
$\sigma$ and temperature in units of $\epsilon/k_B$.

In our calculations we have truncated the Lennard-Jones potential at a 
distance of 3$\sigma$, and shifted the value of the potential within 
this cutoff distance 
by a constant so that the resulting potential approaches zero at 
3$\sigma$ (matching 
the values beyond 3$\sigma$). This truncation is justified because the 
Lennard-Jones potential is already quite small (-0.005$\epsilon$) at this 
distance, and is not expected to significantly affect the accuracy of our 
simulations. Additionally, by using a cutoff distance, we are able to use 
a cell list structure in our algorithms so that our computations scale as 
$O(N)$ instead of the $O(N^{2})$ scaling without a cutoff distance 
($N$ is the number of particles in our simulation cell).

We have collected data for systems of 1600, 6400, and 
25600 particles over a wide temperature range at a density of 
0.873. Additionally, we have simulated a system of 102400 particles 
for three temperature values at the same density in order to verify our 
results for the smaller system sizes. To accommodate the expected 
low temperature triangular solid phase, a periodic simulation cell 
of proportion 2:$\sqrt{2}$ is used. 
We have performed our calculations  on the Florida State University shared
High-Performance Computing facility, which contains several thousand compute nodes.
The  processors on these nodes range  in speed from 2.3 GHz to 2.8 GHz,
and it takes about 34 hours to perform 1,000,000 Monte
Carlo sweeps for  $N=25600$ particles, including calculating observables
every 100 Monte Carlo sweeps (MCS).
We have found that, except for the $N=102,400$ particle system,
a million MCS are sufficient to reach equilibrium,
even near the critical points.
The data presented here is obtained over one million MCS,
after a period of one (for $N=102,400$)
or two (for $N=1600, 6400$, and 25600) million MCS of equilibration. 

To take advantage of our
  computational  resources,  we utilized  a  trivially parallel  Monte
  Carlo  implementation of 100  threads, each  with a  unique random
  number  seed and  initial configuration.  Simulations begin  from an
  initial  near-ordered configuration  (particles are placed  in a
  triangular    lattice,    with    5\%    lattice    spacing    random
  fluctuations). Statistics for  thermodynamic variables are collected
  by generating  averages on  each of the  100 parallel  threads, then
  using the central  limit theorem we obtain the total average, as we have  100 independent means.
 
 Although in  our preliminary studies we  have computed thermodynamic
  quantities for a range of densities and temperatures, the effects of
  critical slowing down near the  melting transition and our desire to
  study the largest possible systems have  led us to focus on a single
  density,  0.873 (all  densities  are  in units  of
  particles per $\sigma^{-2}$). This density was chosen for several 
reasons. This
  is  a density  that could  be  readily compared  to prior  numerical
  simulations of Lennard-Jones melting \cite{Udink}. Also, we wanted a density
  that is  relatively low, but  large enough to avoid  the solid-vapor
  coexistence phase at low temperatures. Strictly speaking, there is a
  solid phase  in the  zero temperature limit  only at densities  of 
  0.9165 (the density  at which the spacing of  the triangular lattice
  is the same as the  position of the Lennard-Jones potential minimum)
  and  above. Below this  density there  is a  solid-vapor coexistence
  phase. However, the triple point  density is roughly 0.82, so at higher
  densities the system will in  general become solid before the  onset of melting
  occurs \cite{Chester}. 


\section{Role of defects}
\label{defects}

\subsection{Defect types}

In two dimensions, the densest packing of particles of uniform size is
achieved  in  a triangular  lattice.  In  such  a configuration,  each
particle has exactly six  nearest neighbors.  Thermal fluctuations will lead to  distortions in the lattice, or even
destroy  it  completely.  To   quantify  this,  we  use  the  Delaunay
triangulation  to  determine  the  nearest  neighbor  network  of  our
particle  configurations. The  nearest neighbor  network tells  us the
number  of   nearest  neighbors,  or  coordination   number,  of  each
particle. For a  system of particles in a  periodic plane, the average
coordination number  is always six \cite{Allen}.  Particles  in a triangularly
ordered  region  will be  six-coordinated,  while  disruptions in  the
lattice will lead to  particles with coordination numbers greater than
or  less than six.   A defect  is defined  as any  coordination number
other than six.  These non-six-coordinated atoms may be  thought of as
disclinations of charge ${\bf n}$, their coordination number being 
$6+{\bf n}$.

The  most  common  type  of  disruption,  or defect,  is  a  five-  or
seven-coordinated particle.  These may be interpreted as disclinations
of  charge plus  or  minus one. 
Two oppositely  charged disclinations may be thought  of as a
dislocation. More complex
arrangements of disclinations are  possible, such as dislocation pairs
and grain boundary loops, but  in our analysis we have only considered
individual defects. The defect fraction, $f_{d}=1 - N_{6} / N$, is defined as
the fraction of  particles that do not have six  neighbors, where $N$ is
the  number of  particles  in the  system,  and $N_6$  is  the number  of
six-coordinated   particles   in   the   system.    Remembering   that
dislocations are  made of two bound disclinations  of opposite charge,
and that dislocations  become unbound above the melting  point, we can
expect the defect fraction to experience a jump at the melting point
\cite{Allen}. Additionally, at low temperatures we can expect an energy gap to
occur, which is the energy cost to create a dislocation pair. Because
the  overall disclinicity of  the system  must be zero,
as  well  as   the  net  Burgers  vector  of   any  dislocations,  the
lowest-energy  defect excitation is  a dislocation  pair of  opposite Burgers
vectors.   In   practice  this  is   usually  two  pairs  of   5-  and
7-coordinated particles. This leads  to an exponential behavior in  the defect fraction,
$f_d  = e^{-\beta \Delta}$,  where  $\Delta$  is the  lowest-energy
for a defect type excitation  of  the system.

\subsection{Unbinding of defects}

\begin{figure}
\vskip 0.3 in
\begin{center}
\includegraphics[width=\figwidth]{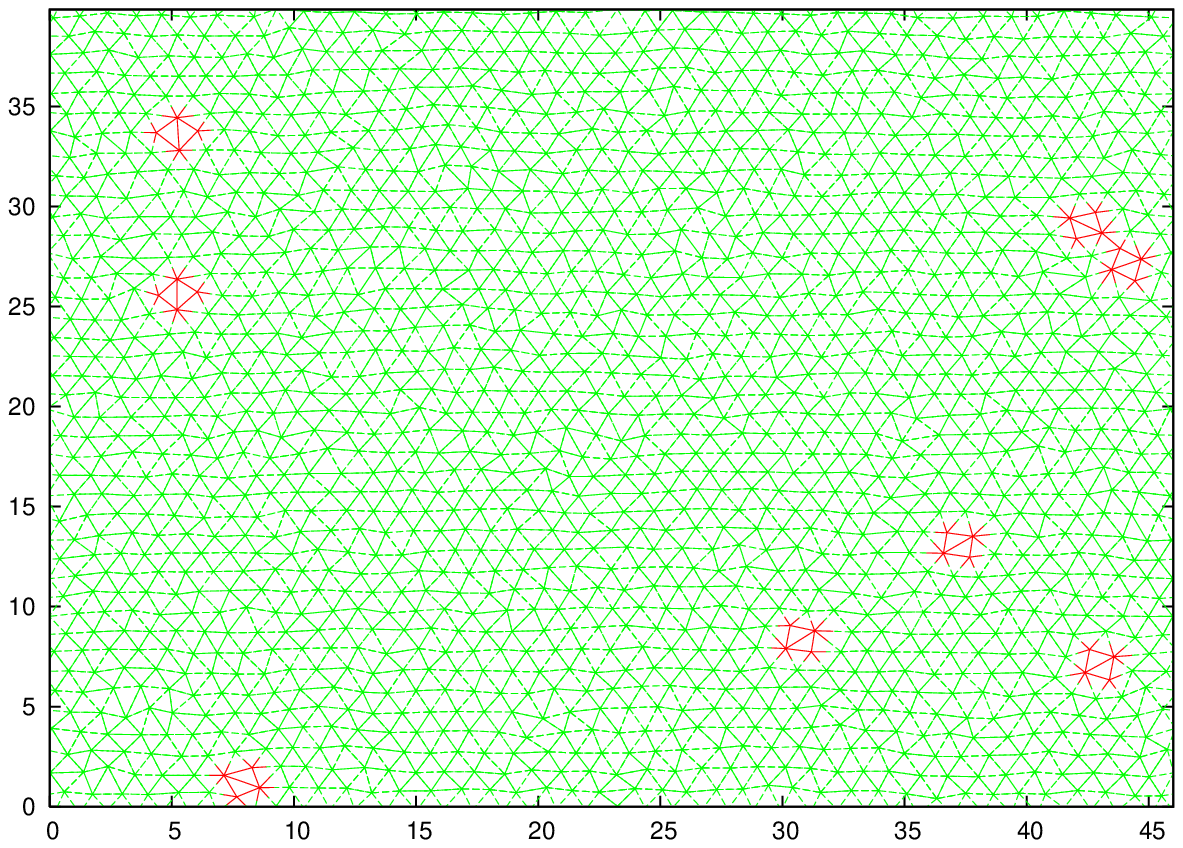}
\end{center}
\caption{The Delaunay triangulation for $N=1600$ particles at
T=0.70. Defects are shown in red.}
\label{data_net070}
\vskip 0.3 in
\end{figure}

\begin{figure}
\begin{center}
\includegraphics[width=\figwidth]{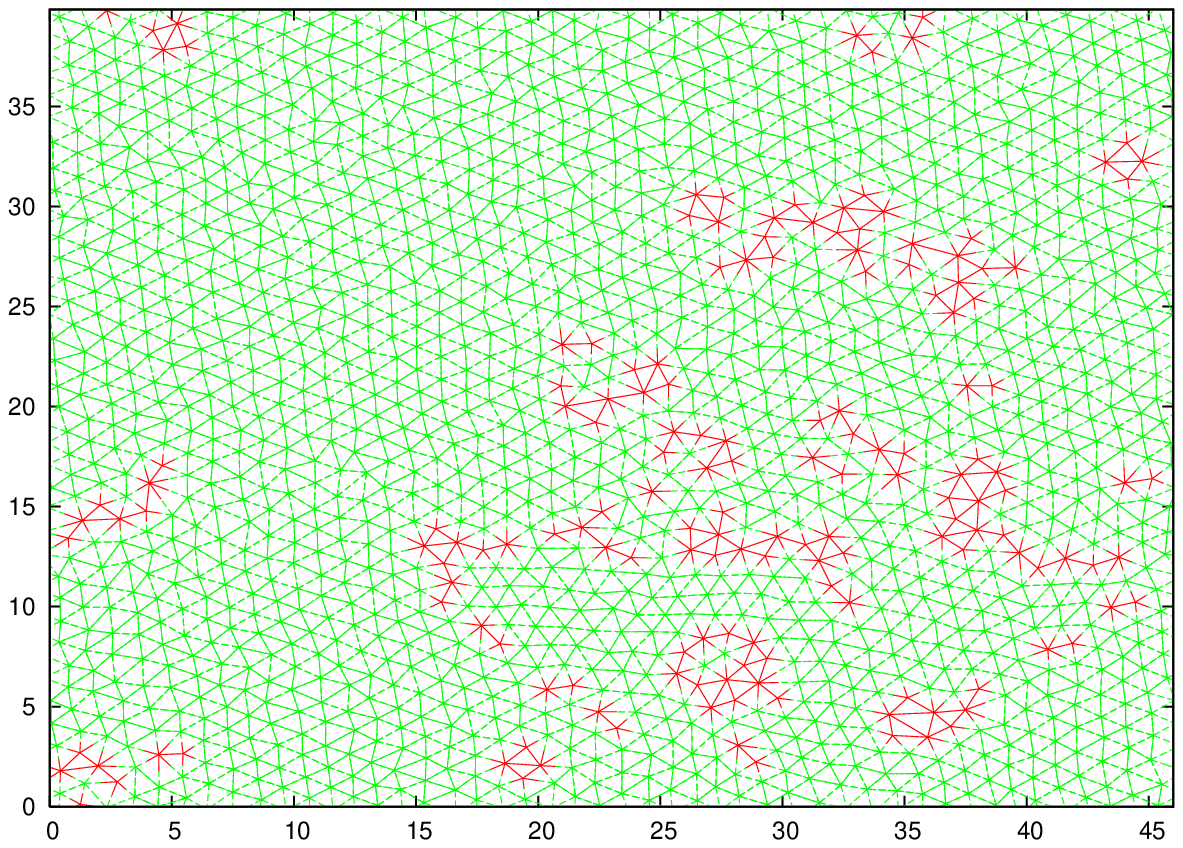}
\end{center}
\caption{The Delaunay triangulation for $N=1600$ particles at
T=0.90. Defects are shown in red.}
\label{data_net090}
\end{figure}

\begin{figure}
\vskip 0.3 in
\begin{center}
\includegraphics[width=\figwidth]{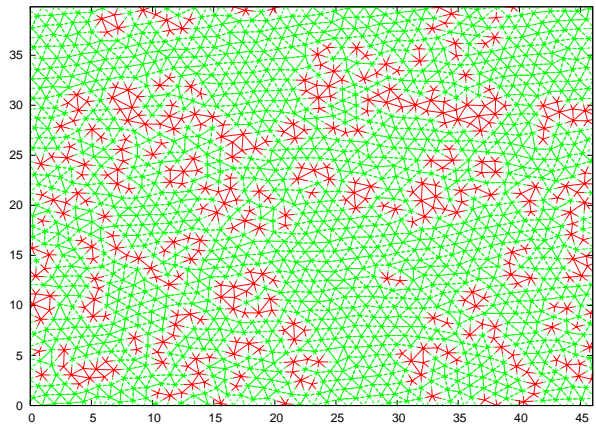}
\end{center}
\caption{The Delaunay triangulation for $N=1600$ particles at
T=1.10. Defects are shown in red.}
\label{data_net110}
\vskip 0.3 in
\end{figure}

In Figures~\ref{data_net070},\ref{data_net090},\ref{data_net110} the
Delaunay triangulated configuration of a 1600 particle system is shown
at temperatures 0.7, 0.9 and 1.1 respectively. The defects are shown in red.  
At low temperature as demonstrated in Figure~\ref{data_net070},  we  see that  
defects occur  in  quadruplets  consisting of two  5-coordinated and  
two 7-coordinated  particles. As the  temperature is raised to 0.9
(Figure~\ref{data_net090}) we can see isolated dislocations (one
5-fold coordinated atom bound to a 7-fold coordinated atom). At yet
higher temperature, such as 1.1 (Figure~\ref{data_net110}) we can
observe isolated disclinations.

\begin{figure}
\vskip 0.3 in
\includegraphics[width=\figwidth]{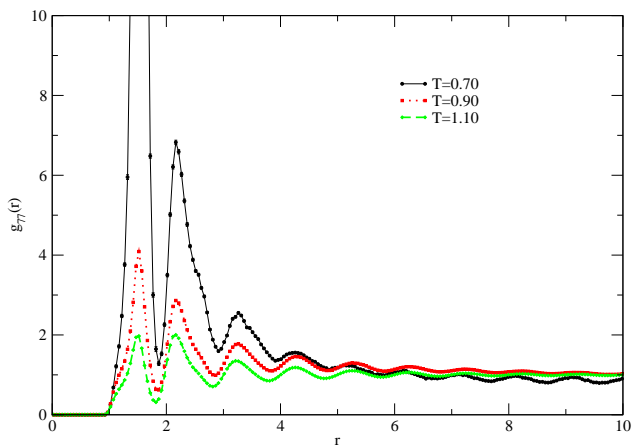}
\caption{The pair distribution function for 7-coordinated particles,
$g_{77}(r)$. The peak for T=0.70 extends to $\sim 50$.}
\label{data_g77}
\vskip 0.3 in
\end{figure}

This can also be seen in 
the pair distribution  functions $g_{77}(r)$, $g_{55}(r)$, and $g_{57}(r)$, 
for pairs of 7-coordinated particles, pairs of 5-fold coordinated atoms and for
5-fold-7-fold coordinated atoms respectively. 
In Figure~\ref{data_g77}, 
a sharp peak in  $g_{77}(r)$  is observed  at  low 
temperatures ($T=0.70$), indicating  that dislocations are tightly  bound. 
At higher temperatures ($T=0.90$ and $T=1.10$), the peak in $g_{77}(r)$ is 
greatly diminished, and dislocations become first weakly bound ($T=0.90$) 
and then completely unbound ($T=1.10$). $g_{55}(r)$, while not shown, 
behaves qualitatively similar to $g_{77}(r)$, as both are representative 
of the pair distribution of dislocations.

\begin{figure}
\vskip 0.3 in
\includegraphics[width=\figwidth]{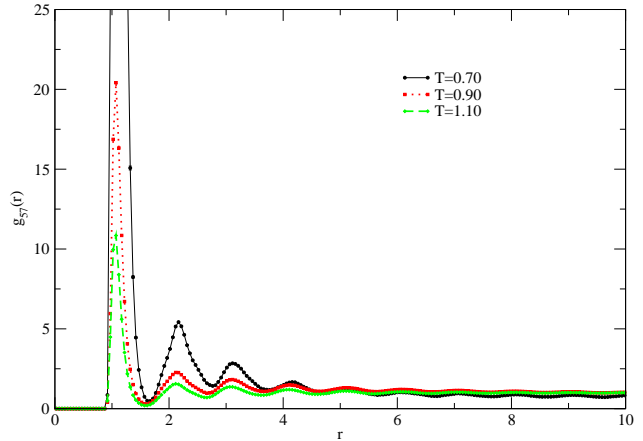}
\caption{The pair distribution function for pairs consisting of
one 5-coordinated particle and one 7-coordinated particle,
$g_{57}(r)$. The peak for T=0.70 extends to $\sim 150$.}
\label{data_g57}
\vskip 0.3 in
\end{figure}

The pair distribution  function for
disclinations, $g_{57}(r)$, is  shown in Figure~\ref{data_g57}. 
While  the sharp peak
at low ($T=0.70$) and intermediate ($T=0.90$) temperature is expected, the
peak at  $T=1.10$, while  quite lower, is  still very  substantial. This
indicates that disclinations have not become completely unbound, 
and indeed it is difficult to find isolated disclinations in the snapshot
configurations presented  in Figure~\ref{data_net110}.  
When isolated disclinations
do  occur, they  are still  next-nearest neighbors  with at  least one
other disclination of opposite charge.

\subsection{Defect fraction}

\begin{figure}
\vskip 0.3 in
\includegraphics[width=\figwidth]{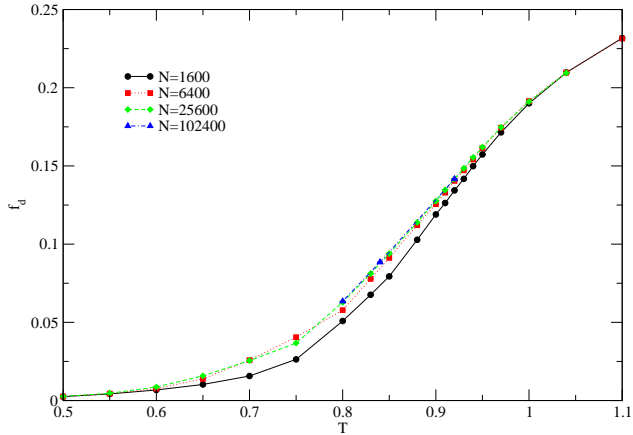}
\caption{Fraction of defects, $f_d$, as defined by the fraction of
non-six-coordinated particles in the Delaunay triangulation, $f_d=1-N_6/N$.
The rapid rise in $f_d$ from near zero to almost 25\% is a possible
sign that dislocation and/or disclination unbinding is occurring.}
\label{data_defects}
\vskip 0.3 in
\end{figure}


\begin{figure}
\vskip 0.3 in
\includegraphics[width=\figwidth]{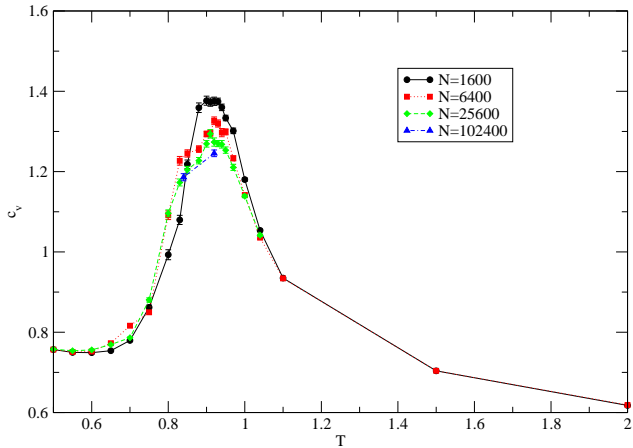}
\caption{The presence of a peak in the specific heat is indicative
of a phase transition.
Interestingly, the peak near $T=0.9$ appears to
lessen in magnitude as the system size is increased.}
\label{data_specific_heat}
\vskip 0.3 in
\end{figure}

\begin{figure}
\vskip 0.3 in
\includegraphics[width=\figwidth]{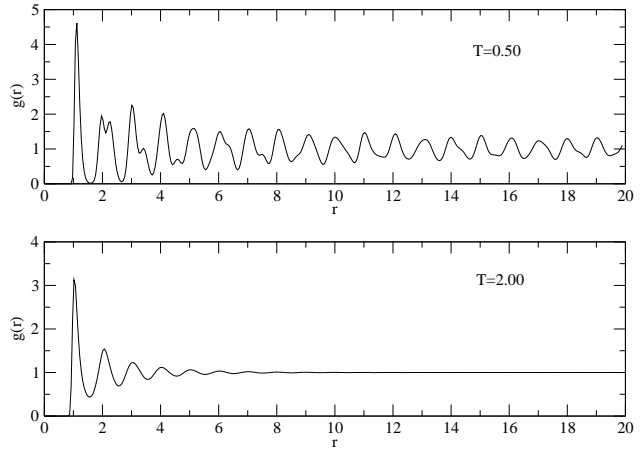}
\caption{The distribution function shows ordering at low temperatures,
as shown above for $T=0.50$, while at higher temperatures, such as T=2.00
shown above, there is a loss of order over moderate length scales.}
\label{data_distrib}
\vskip 0.3 in
\end{figure}

According to the KTHNY
  theory, disclinations remain very tightly bound below $T_m$.  Above $T_m$,
  the disclinations are  screened from one another by  the presence of
  free dislocations  yet remain bound, albeit by  a weaker logarithmic
  binding\cite{NH}. Thus,  we expect  a proliferation  of defects  to occur
  around $T_m$, and to continue growing until somewhere above $T_i$, where a
  saturation should  occur. In Figure~\ref{data_defects} we show the  
average defect
  fraction as a function of temperature. At low temperature, there are
  very few defects, while at  high temperature there is a considerable
  fraction  of the system  that is  defected. In  between, there  is a
  region of  rapidly increasing defect fraction,  from $T = 0.8$  to $T =
  1.0$. 
This  can be  quantitatively verified  by  calculating the temperature 
derivative of the defect fraction, which is indeed found to have a 
broad peak in this temperature region. The overall shape of $df_d(T)/dT$ 
is very similar to that of the specific heat capacity, 
to be shown next.
 Additionally, we  can see some size  dependence in the 
 region $0.6 < T  < 1.0$,  although this seems to be an issue mostly
 for comparisons of  the smallest system size ($N=1600$)  to the larger 
 system sizes. 

The specific heat per particle at constant volume, $c_V$, 
can be calculated from the energy fluctuations,
\begin{eqnarray}
c_V = {1 \over N} {{\langle E^2 \rangle - \langle E \rangle^2 } \over {k_B T^2}}
\end{eqnarray}
where  $E$  is the  total  energy  of an  $N$  particle  system.  We  have
calculated the specific heat and  show it as a function of temperature
in  Figure~\ref{data_specific_heat}. 
One  can see  a broad  peak in  the specific  heat per
particle. According to the KTHNY  theory, there should be an essential
singularity in the specific heat at both $T_m$ and $T_i$\cite{NH}. However, it 
is not clear whether this  will be visible above background contributions
to the specific heat. Either way,  the peak in specific heat points to
a rearrangement of order in the  systems studied.  Also, if we look at
the  distribution  function  (Figure~\ref{data_distrib}),  
we see  ordering  at  low
temperatures, and  fluid behavior at high temperatures.  Overall, it is
clear that  there is a  phase transition occurring, with  a disordered
fluid  state  at  high  temperatures  and  an  ordered  state  at  low
temperatures. 
%
%
%

\subsection{Defect excitation energy}

In the KTHNY theory, dislocations are bound
  at low  temperatures, and there  is a defect core  energy associated
  with their creation. This leads to an energy gap, and thus using the
  Arrhenius law,  we expect $f_d  = e^{-2E_c/k_BT}$,  where we have  used 
$2E_c$  because  dislocation pairs   are  the  lowest  energy  excitation
  (isolated  dislocations are  forbidden). In  Table~\ref{table1}  
we  show the
  defect activation energy  as calculated by the Arrhenius  law at low
  temperatures.  Taking the low temperature limit, we find $E_c = 1.49\pm
  0.01$.  

\begin{table}[hb]
\begin{tabular}{|p{50pt}|p{50pt}|p{50pt}|p{50pt}|} \hline
 Temperature & N=1600 & N=6400 & N=25600 \\ \hline
0.50 & 1.49946(30) & 1.4919(19) &  1.4873(14) \\ \hline
0.55 & 1.50267(85) & 1.4884(31) &  1.4778(22) \\ \hline
0.60 & 1.4996(14)  & 1.4635(43) &  1.4252(31) \\ \hline
0.65 & 1.4872(26)  & 1.3908(59) &  1.3480(29)  \\ \hline
0.70 & 1.4543(30)  & 1.2795(49) &  1.2835(15) \\ \hline 
0.75 & 1.3640(54)  & 1.2027(19) &  1.2390(26) \\ \hline
\end{tabular}
\caption{Defect  activation energy  for
  various  system  sizes  and  temperatures,  as  computed  using  the
  Arrhenius law. The numbers in parentheses are the uncertainty of the
  trailing digits. }
\label{table1}
\end{table}

\section{Order parameters}
\label{orders}

\subsection{Definition and temperature dependence}

Let us define a global order parameter of translational order,
\begin{equation}
\Psi_{\vec{G}}=\frac{1}{N}\sum_{j=1}^{N}\exp\left(i\vec{G}\cdot\vec{r_{j}}\right),
\end{equation}
where $\vec{G}$ is a reciprocal lattice vector, and $\vec{r_{j}}$ is
the position vector of particle $j$. If there is translational
ordering in a system, then clearly $\Psi_{\vec{G}}$ will be non-zero
if $\vec{G}$ is a reciprocal lattice vector of the appropriate lattice
geometry. Due to the shape of our simulational cell, at low
temperatures this will be a triangular lattice with nearest-neighbors
in the x-direction. At high temperatures, no translational ordering is
present, and all possible values of $\vec{G}$ should give the same
(qualitative) result. However, at intermediate temperatures, it may be
possible for there to be some degree of translational ordering that is
not strictly commensurate with our simulation cell. Indeed, we have
observed ``canted'' solid phases at intermediate temperatures, where
we find partial triangular order with nearest neighbors in a direction
titled from the x-axis by a small angle. In this case, if $\vec{G}$
for the triangular order commensurate with our simulation cell is
used, $\Psi_{\vec{G}}$ will be found to be zero. However, if we use an
appropriate $\vec{G}$ for the order present, $\Psi_{\vec{G}}$ will be
found to be non-zero. For this reason, we define the true
translational order to be the maximum value of $\Psi_{\vec{G}}$ for
all $\vec{G}$. In practice, it is not possible to perform this
optimization for each Monte Carlo configuration, so we make the
following assumptions. First, due to the nature of ordering in two
dimensions, we assume any lattice will be triangular. Second, because
the density of particles is fixed, we assume the lattice spacing in
said triangular solid to be the same as that for the commensurate
cell. Thus, we keep the magnitude of $\vec{G}$ constant, and simply
determine the direction of solid ordering for each configuration by
looking at the average bond direction between nearest neighbor
particles. This turns out to be a good estimate of the true
translational order for a system, but it must be remembered that it is
strictly speaking a {\em lower bound}.

 The local order parameter which
measures the degree of six-fold orientational ordering is defined as
\begin{eqnarray}
\psi_6(\vec r_i) = {1 \over {n(i)}} \sum_{j=1}^{n(i)} e^{i 6
  \theta_{ij}}\
\end{eqnarray}
where $\theta_{ij}$ is the angle of the bond between particles $i$ and
$j$ and the sum over $j$ extends over all $n(i)$ nearest neighboring atoms
found by the Delaunay triangulation. 
The global order parameter associated with bond-orientational
order is obtained as an average over all particles.
\begin{eqnarray}
\Psi_6 = {1 \over N} \sum_{i=1}^N \psi_6(\vec r_i).
\end{eqnarray}
In a perfectly bond-ordered triangular solid, we have that
$n(i)=6$ and $\theta_{ij}=\pi/3$ for all $j=1,..,6$. In such case 
$|\langle \Psi_6 \rangle| =1$. In the low temperature phase, there is
bond-orientational order, so $\langle \Psi_6 \rangle$ should be a
point on the perimeter of a circle with a radius approaching unity as
$T \to 0$. In the hexatic phase there is quasi-long-range
bond-orientational order, which implies that the distribution of
$\langle \Psi_6 \rangle$ should become a ring in the imaginary
plane. In the isotropic phase both $\langle \Psi_6 \rangle$  and 
$\langle \Psi_{\vec G} \rangle$  should be distributed around zero value. 

\begin{figure}
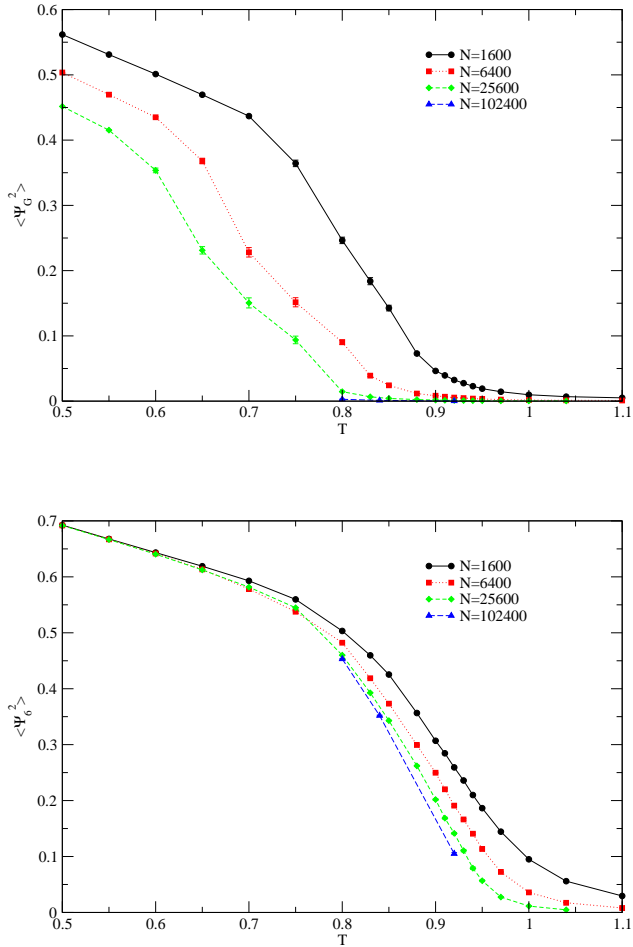

\vskip 0.3 in
\begin{center}
\includegraphics[width=\figwidth]{Fig9a.eps}
\vskip 0.4 in
\includegraphics[width=\figwidth]{Fig9b.eps}
\end{center}
\caption{The second moment of the translational (top) and
bond orientational (bottom) order parameters.}
\label{data_psi2}
\vskip 0.3 in
\end{figure}

In  the top  panel of  Figure~\ref{data_psi2} we show  the second  
moment  of the
translational  order parameter, $\Psi_G^2$  .  There  appears to  be a
transition from a translationally ordered phase at low temperatures to
a disordered phase at higher  temperatures. In the ordered phase there
is a  clear relation  between $\Psi_{\vec G}^2$  and system  size.  We
will explore  this relation  in a  later section, but  for now  let us
point out that this finite-size  scaling relation begins to break down
above $T=0.60$. This is expected within the KTHNY theory of melting
due to the unbinding of dislocations. However, on closer inspection,
 the behavior of the curves
for $N=6400$ and $N=25600$ in the region $0.6<T<0.8$  is not a smooth
connection of the data at higher and lower temperature. This is due to our
measured quantity $\Psi_{G}$ being a lower bound of translational order.

Also  shown  in   Figure~\ref{data_psi2} is  the  second   moment  of  the  bond
orientational  order  parameter,  $\Psi_6^2$  (bottom panel).  At  low
temperatures there is substantial bond orientational order.  Below $T =
0.70$ there is very little dependence of $\Psi_6^2$ on system size.  As
the  temperature is  increased,  $\Psi_6^2$ begins  to  show a  marked
dependence on  system size as well as  a steep decline in  value as we
approach the high temperature disordered phase.

\subsection{Order parameter distribution}

\begin{figure}
\vskip 0.3 in
\begin{center}
\includegraphics[width=\figwidth]{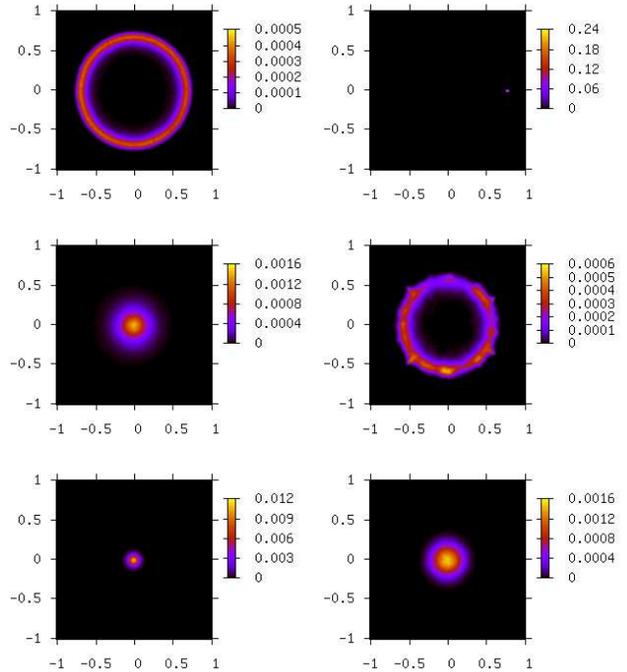}
\end{center}
\caption{Intensity plots of the probability of $\Psi_{\vec{G}}$ (left column)
and $\Psi_{6}$ (right column) on the complex plane for
T=0.70 (top row), T=0.90 (middle row), and T=1.10 (bottom row).}
\label{data_psimaginary}
\vskip 0.3 in
\end{figure}

The main prediction  of Halperin and Nelson\cite{HN,NH}  and 
Young\cite{Young} is
that if two dimensional melting is the result of dislocation unbinding,
as proposed  by Kosterlitz and  Thouless\cite{KT}, then a  second unbinding
transition (of disclinations) is  required to reach an isotropic fluid
state. This  implies the presence of  a novel hexatic  fluid phase. In
Figure~\ref{data_psimaginary} we  show an intensity plot of the
distribution of $\Psi_{\vec G}$ and of $\Psi_6$  
on the complex plane
for three different temperatures.

At $T=0.70$ (top row), our calculation of the distribution of the
order parameters finds a ring
of values for  $\Psi_{\vec G}$, while $\Psi_6$ is  localized in a small
region away from  the origin (a very narrow peak showing as a ``star''
along the positive real axis). This is consistent  with the presence of
long-range bond orientational order ($|\langle\Psi_6\rangle | >0$), 
while the ring of $\Psi_{\vec G}$ values is expected for quasi-long-range 
translational order. At $T=0.90$ (middle  row), 
we that $\Psi_{\vec  G}$ is  clustered  about   the    origin,  
indicating   a    lack   of   translational  order. 
Interestingly, $\Psi_6$ now shows  a ring of values about the 
origin, indicating  quasi-long-range order. This is  exactly what is 
expected of the hexatic fluid phase. Finally, at $T=1.10$ (bottom row) 
we see that both order  parameters are distributed about the origin, 
indicating an isotropic  fluid phase of no order.   

\section{Critical exponents}
\label{exponents}

In the topological solid phase, the scaling form for the second moment 
of the translational order parameter is $\langle \Psi_G^2\rangle \sim L^{-\eta}$,
where $L$ is the (linear) system size and $\eta$ is a critical
exponent. In the hexatic fluid phase, a similar relation holds for bond 
orientational order, $\langle \Psi_6^2 \rangle \sim L^{-\eta_6}$.
According to the KTHNY theory of melting, the critical exponents 
$\eta$ and $\eta_6$ will have
specific values at melting. The translational critical exponent 
is bounded at lower melting temperature: $1/4 < \eta(T_m) < 1/3$. 
Additionally, the bond orientational critical exponent grows
from zero at $T_m$ to 1/4 at $T_i$, and is related to the translational 
correlation length: $\eta_6(T) \sim \xi(T)^{-2}$ \cite{NH}. 

\begin{figure}
\vskip 0.3 in
\begin{center}
\includegraphics[width=\figwidth]{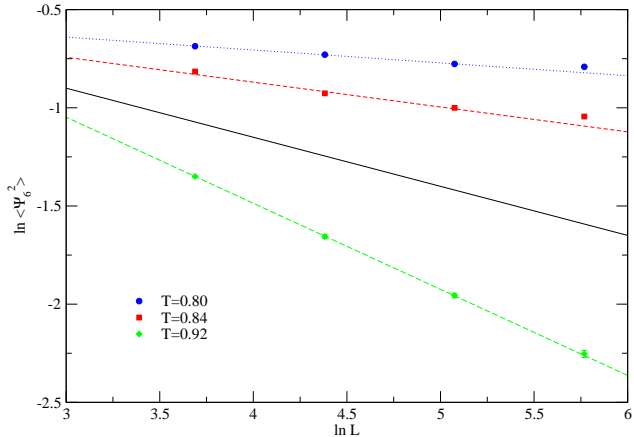}
\end{center}
\caption{Scaling of $<\Psi_6^2>$ with system size $L$,
shown on a logarithmic plot.
Results for T=0.84 are shown as blue circles,
while data collected at T=0.92 is represented by red squares.
In both cases, the data for the three smaller systems was
fit to the equation $\ln<\Psi_6^2>=-\eta_6\ln L+const$,
and the result is plotted as the dotted and dashed lines
(the solid line is the KTHNY value of $\eta_6$ at $T_i$).
In both cases, the value of
$\ln<\Psi_6^2>$ of the largest system size ($N=102400$) is
reasonably close to the value expected from scaling.}
\label{data_logeta6}
\vskip 0.3 in
\end{figure}

\begin{figure}
\vskip 0.3 in
\begin{center}
\includegraphics[width=\figwidth]{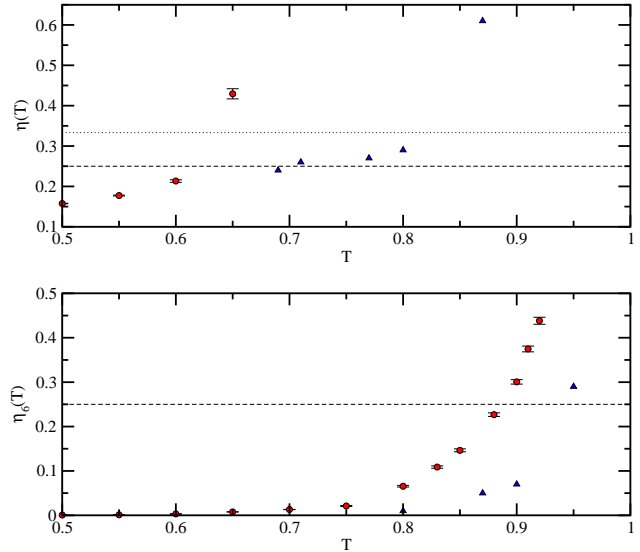}
\end{center}
\caption{Anomalous dimensionality of (top) the translational order
parameter and (bottom) the bond orientational order parameter.
Our current results are shown in both figures as red circles.
Shown for comparison are the results of Udink and van der Elsken~\cite{Udink} (blue triangles, both figures).
In the top figure, the dashed and dotted lines represent the lower and upper
bounds of $\eta$ at $T_m$, according to KTHNY theory; in the bottom figure,
the dashed line represents the predicted value of $\eta_6$ at $T_i$.}
\label{data_etas}
\vskip 0.3 in
\end{figure}

By plotting  $\langle \Psi_G^2\rangle$ (or $\langle \Psi_6^2\rangle$)
versus $L$ on a log-log plot, we can find $\eta$ (or $\eta_6$).
 To  demonstrate the validity of this scaling law and  that  our
 results are  not  limited  by system size, in
 Figure~\ref{data_logeta6}  we plot $\langle \Psi_6^2 \rangle$ versus
 $\ln L$  for all system  sizes at  the two  temperatures where  we
 have results for the $N=102400$ system. 
Results of linear least squares fits
to  the three smallest  system sizes  (used to  generate the  data for
Figure~\ref{data_etas}) are  shown as a dotted blue 
line (for  data at T=0.80), a
dashed red  line (for  data at T=0.84),  and a long-dashed  green line
(for data at  T=0.92).  For the higher temperature,  the $N=102400$ data
falls directly  on this line,  within error bars. At  T=0.84, however,
the  $N=102400$ data  indicates  that a  smaller  value for  $\eta_{6}$ 
at  this temperature may be necessary. This could either be due to the 
(presumably)  large translational   correlation   lengths  at   this
temperature, which would invalidate results for small system sizes, or
perhaps a very  long relaxation time. Either way, from  our data it is
clear that by T=0.92 the KTHNY value of $\eta_6$ at $T_i$ has been well passed.

In Figure~\ref{data_etas} we show the  extracted values of  $\eta$ and $\eta_6$,
the critical exponents of  translational and bond orientational order.
In both panels, we show our  results as red circles. In the top panel,
we  can  see  that $\eta$  crosses  the  KTHNY  melting value  in  the
temperature range  $0.6 <  T < 0.65$.  In the  bottom panel we  show the
critical exponent of bond orientational order, $\eta_6$. This exponent
crosses the  KTHNY melting  value (see dashed  line) at  a temperature
near 0.89,  in close agreement with  the value for  $T_i$ derived from
the  divergence of  the  correlation length $\xi_6$ obtained in the next section. 
This value for $T_i$ is also in good agreement with the value
reported in Ref.~\onlinecite{Keola10}.
However, in
Figure~\ref{data_etas} we also show the algebraic  exponents reported by 
Udink and van der
Elsken\cite{Udink}.  In both panels, we  can see that their  values cross the
KTHNY melting zone at higher temperatures than our values.  We believe
this disagreement  may be due  to insufficient thermalization  time in
their  study, as this  could lead  to artificially  low values  of the
critical  exponents.
We should sound a note of caution here in regards to the scaling of $\langle \Psi_G^2\rangle$.
Because our measurements for $\Psi_{G}^{2}$ are lower bounds, it is possible that the extracted
exponents $\eta(T)$ are not correct in the temperature regime where $\vec{G}$ is no longer
commensurate with the simulation cell, as is the case for $T>0.6$.

\section{Correlation Functions}
\label{correlations}

\begin{figure}
\vskip 0.3 in
\includegraphics[width=\figwidth]{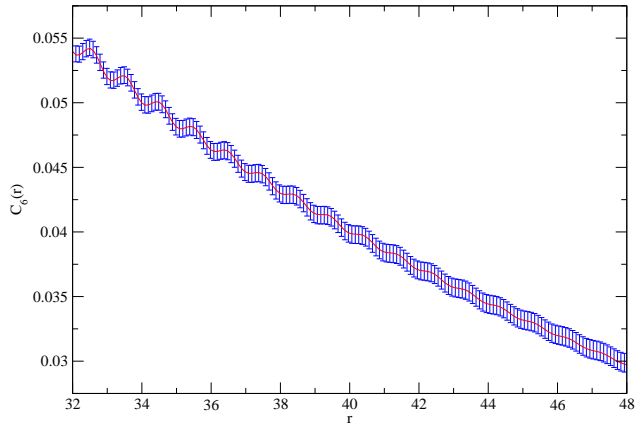}
\caption{Example of fitting the bond orientational correlation
function, $C_{6}$, to the form shown in Equation~\ref{data_eq_g6_fit}.
The data is for $N=25600$ particles at $T=0.97$. The critical
exponents are fixed at their maximum values, $\eta=0.33$ and
$\eta_{6}=0.25$. The extracted correlation lengths are
$\xi=7.40\pm0.19$ and $\xi_{6}=32.6\pm0.7$.}
\label{data_g6_fit097}
\vskip 0.3 in
\end{figure}

The correlation function for bond orientational order is given by
\begin{equation}
C_{6}(r)=<\psi_{6}(r)\psi_{6}^{*}(0)>,
\end{equation}
where $\psi_6(\vec r)$ is the local bond-orientational order
parameter defined in Sec.~\ref{orders}. 
In the isotropic fluid phase, 
the asymptotic form of $C_6(r)$ is  $\sim \exp (r/\xi_6)$ \cite{NH}. 
At shorter distances, however, a
power law decay comes into play, such that as $\xi_6$ diverges as $T_i$ 
is approached from above,
%
%
then the asymptotic form becomes $C_6(r) \sim r^{-\eta_6}$ at $T_i$ and below,
with $\eta_6(T_i) = 1/4$.
Additionally, we observe
oscillations in $C_6(r)$ that 
seem to decay with an exponential envelope. Thus, we used the 
following fitting form
for the bond orientational correlation function for distances $r$ 
much less than the system size $L$:
\begin{eqnarray}
C_6(r) = A {{e^{-r/\xi_6}} \over {r^{\eta_6}}} + B
\sin (kr+\delta){ {e^{-r/\xi}} \over {r^{\eta}}}.
\label{data_eq_g6_fit}
\end{eqnarray}
We have  used a particle-centric definition of  the bond orientational
correlation function, so in our calculations of $C_6(r)$ there will be an
influence from $g(r)$,  the pair distribution function. In  the limit of
perfect  bond orientational ordering  ($\psi_6 =  1$ everywhere),  
$C_6(r)$ and $g(r)$ will  be equivalent.  We approximate  the oscillatory portion  
of $C_6(r)$ which is due to the translational atomic arrangement using a
damped oscillator.  The periodic form is captured by using 
$\sin (kr+\delta)$,
where k is expected to be  near the first reciprocal lattice vector in
magnitude ($\sim 6$) and $\delta$ is just a phase-shift parameter. The size of the
  oscillations are expected to decay exponentially in the fluid phase,
  and algebraically in the hexatic phase,  so we add also a power law,
  ending up with a term $sin(kr+\delta)r^{-\eta} e^{-r/\xi}$.
An example fit is shown in Figure~\ref{data_g6_fit097}.
Note that the fitting procedure returns parameters much more precise than the error bars
in Figure~\ref{data_g6_fit097} would indicate are possible. This is due to the high degree
of correlation between neighboring points of $C_6(r)$. In fact, up to a separation of $3$
the values of $C_6(r)$ are still 99\% correlated! This simply means that the relative form
(including the rate of decay) of $C_6(r)$ is consistent between our various calculations,
remembering that we average the values of 100 independent parallel Monte Carlo simulations.

We wish to note that we observe an upturn in $C_6(r)$ as $r$ approaches $L/2$. 
At temperatures closer to melting
(larger correlation lengths), the upturn occurs further from $L/2$.
Next, we would like to determine a characteristic distance $R$ for a given
finite-system of linear dimension $L$ so as to stay away from this upturn due to
finite-size effects. Namely, we wish to limit the range of $r$ in our
fit of the correlation function to the form given by Eq.~\ref{data_eq_g6_fit} in the range
$\xi_6<r <R$. Let us assume a periodic
form for the correlation function:
\begin{equation}
C_6(r) = A \left(\frac{exp(-r/\xi_6)}{r^{\eta_6}} + \frac{exp(-(L-r)/\xi_6)}{(L-r)^{\eta_6}} \right)
\end{equation}
Neglecting the power-law term, the upturn is expected to occur
 when the $L - r$ terms are a significant fraction of the $r$ terms. Thus,
\begin{eqnarray}
R= {L \over 2} - {{\xi_6} \over 2} \ln(x)
\end{eqnarray}
 where $R$ is the distance at which the $L - r$ terms are a fraction $x$ 
of the $r$ terms. Using $x = 0.05$ or 5\%, this leads to
\begin{eqnarray}
R= {L \over 2} - {{3 \xi_6} \over 2} .
\end{eqnarray}

\begin{figure}
\vskip 0.3 in
\includegraphics[width=\figwidth]{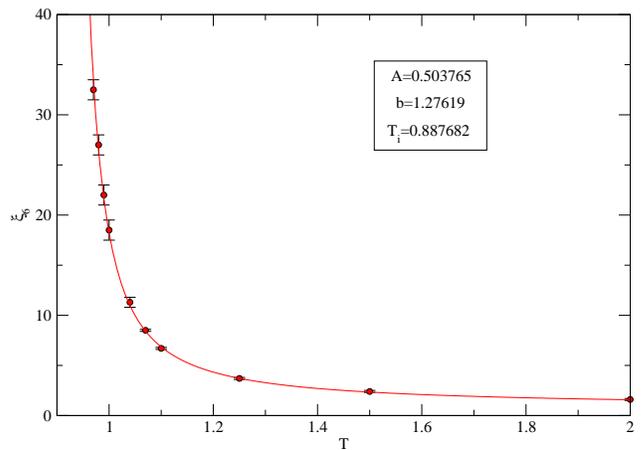}
\caption{Correlation lengths of the bond orientational order parameter
as determined by fitting the bond orientational correlation function to
the form mentioned in the text.}
\label{data_xi6}
\vskip 0.3 in
\end{figure}

In Figure~\ref{data_xi6} we show $\xi_6(T)$ as determined by fitting $C_6(r)$ in the 
range $\xi_6 < r < R$. These values were fit to the KTHNY form of 
the expected divergence of $\xi_6$ as $T_i$ is
approached from above: $\xi_6(T) = A\exp(b/t^{\nu})$, where 
$t = ( T - T_i)/T_i$ and $\nu= 1/2$. This fit gives a value for $T_i$ 
near 0.89.

\begin{figure}
\vskip 0.4 in
\includegraphics[width=\figwidth]{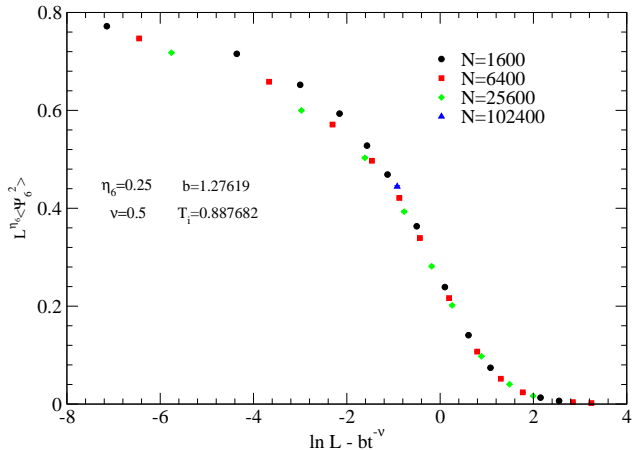}
\caption{Demonstration of finite-size-scaling by plotting the
  dimensionless quantity $L^{\eta_6} \langle \Psi_6^2 \rangle$ versus
  $\ln(L/\xi_6)$, i.e.,   the logarithm of the ratio of the
  finite-system-size to the correlation length, for various size systems.}
\label{fss_bond}
\vskip 0.4 in
\end{figure}

Using the calculated correlation length and critical exponent
$\eta_6$,  in Figure~\ref{fss_bond} we plot the dimensionless quantity 
$L^{\eta_6} \langle \Psi_6^2 \rangle$
as a function of the dimensionless ratio $\ln(L/\xi_6)$ for all
size-lattice considered here. Notice that the data collapse onto the same scaling
function using the same values of the parameters for our fit to $\xi_6(T)$
shown in Figure~\ref{data_xi6}, and also setting $\eta_6=\eta_6(T_i)=1/4$.
This provides additional support for the theory.

\section{Distribution functions}
\label{distributions}

\begin{figure}
\vskip 0.4 in
\includegraphics[width=\figwidth]{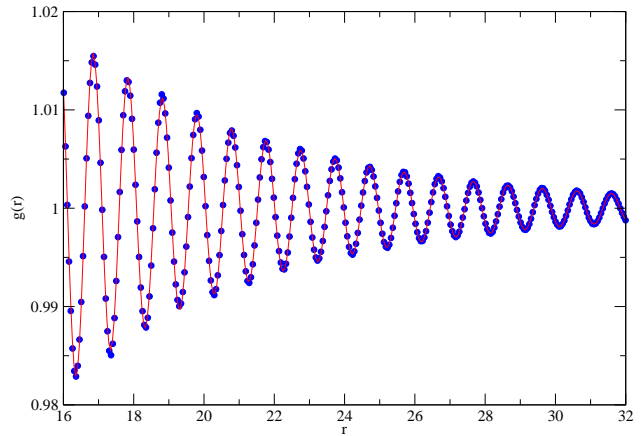}
\caption{Example of fitting the pair distribution function,
$g(r)$, to the form shown in Equation~\ref{data_eq_g_fit}.
The data is for $N=25600$ particles at $T=0.97$. The critical
exponent $\eta$ is fixed at its maximum value, 1/3.
The extracted correlation length is $\xi=8.09\pm0.04$.}
\label{data_g_fit097}
\vskip 0.4 in
\end{figure}


In the disordered phase ($T>T_i$) the distribution function can be
obtained as an angular average of the bond-orientational correlation
function $C_{\vec G} (\vec r)$ as
\begin{eqnarray}
g(r) = 1+ {1 \over {2 \pi}} \int_0^{2 \pi} e^{i\vec{G}\cdot\vec{r}} C_{\vec{G}}(\vec{r})d\phi.
\end{eqnarray}
The integration of $e^{i\vec G\cdot \vec r}$  will give us a zeroth-order 
Bessel function of the first kind, $J_0(Gr)$,
and using the KTHNY form of the translational correlation function, 
$C_{\vec G}(\vec r)  \sim \exp(-r/\xi)r^{-\eta}$,
we wind up with the following form for the radial pair distribution function 
(in the high temperature limit):
\begin{eqnarray}
g(r \to \infty) = 1 + A J_0(Gr) e^{-r/\xi} r^{-\eta}
\end{eqnarray}
where $A $ is some amplitude.

\begin{figure}
\vskip 0.3 in
\includegraphics[width=\figwidth]{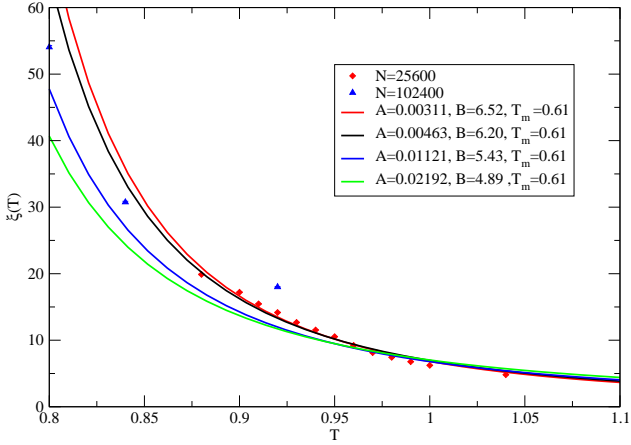}
\caption{Correlation lengths of the translational order parameter
as determined by fitting the pair distribution function to
the form mentioned in the text. The range of the fit is from $2\xi$
to $4\xi$, with $\eta$ fixed at its maximum value of 1/3. At $T=0.80$, the 
fitting range is from $\xi$ to $2\xi$. The solid lines are obtained from the 
 KTHNY form, $\xi = A \exp(b/t^{\nu})$, using $\nu=0.36963$ and various values 
of the other parameters. Our best fit corresponds to the red curve. }
\label{data_xi}
\vskip 0.3 in
\end{figure}

The $\vec G$ that we use here is the same as in the definition of the 
translational order parameter, namely we use the first reciprocal lattice 
vector of the idealized triangular lattice that is commensurate with our 
simulation cell. For the density considered ($\rho\sigma^2 = 0.873$),
this means $G\simeq6.3\sigma^{-2}$. 
Thus $Gr$ is quite large for moderate values of $r$, 
and we can use the asymptotic expansion of $J_0$, 
namely $J_0(x\to\infty) = \sqrt{2/\pi x} \cos(x-\pi/4)$. 
Thus in practice we fit $g(r)$ in the
disordered phase to the following form,
\begin{eqnarray}
g(r\to \infty) = 1 + A\cos(kr + \delta) {{e^{-r/\xi}} \over{ r^{\eta+1/2}}}
\label{data_eq_g_fit}.
\end{eqnarray}
An example  fit is shown in Figure~\ref{data_g_fit097}.  
In Figure~\ref{data_xi}  we show the
correlation  length of  translational order  as calculated  by fitting
$g(r)$ to the above form. Results are shown for the $N=25600$ and $N=102400$
particle  systems.   Clearly,  $\xi$   remains  finite  even   as  the
orientational  correlation length diverges.   However, there  are some
discrepancies in our  values of $\xi$.  At $T=0.92$,  the value of $\xi$
extracted from  the $N=25600$  particle system does  not agree  with the
value  for  $N=102400$  particles.  

We believe that some of this  difference may be
attributable to finite size  effects.  Additionally, there is also the
possibility  that   the  $N=102400$   particle  system  has   not  fully
thermalized.  While  we have  tried to ensure  that the data  for this
largest system is completely thermalized,  it can be very difficult to
distinguish between  stable and metastable states. In  either case, we
can not consistently fit all the data to the KTHNY form, $\xi = A
\exp(b/t^{\nu})$, so instead we have made the fit for only the $N=25600$
data.  The result of a fit with $T_m = 0.61$, $A=0.00311$ and $B=6.62$ using
$\nu=0.36963$ is shown
in Figure~\ref{data_xi} as the red curve. In addition,  a few other 
curves are also shown for different values of these parameters
with the same value of $T_m=0.61$ whose significance is discussed next. 

\begin{figure}
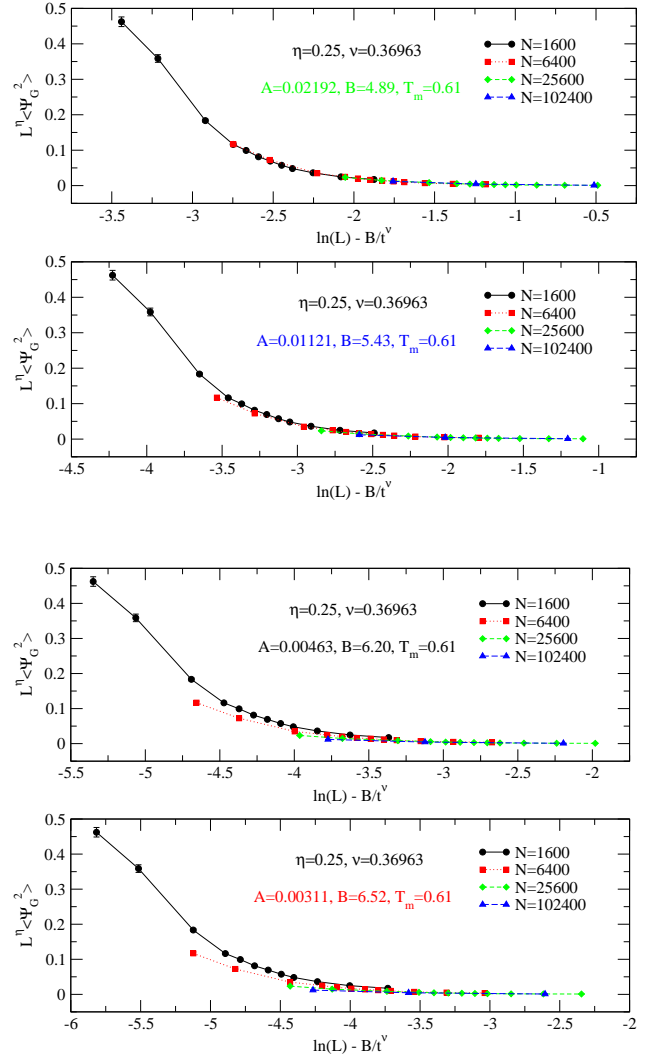

\begin{center}
\vskip 0.3 in
\includegraphics[width=\figwidth]{Fig18a.eps}
\vskip 0.3 in 
\includegraphics[width=\figwidth]{Fig18b.eps}
\caption{Demonstration of finite-size-scaling by plotting the
  dimensionless quantity $L^{\eta} \langle \Psi_{G}^2 \rangle$ versus
  $\ln(L/\xi)$, i.e.,   the logarithm of the ratio of the
  finite-system-size to the measured correlation length, for the two size
  systems, using $\eta=1/4$ and for various parameters.
  Due to our calculation of $\langle\Psi_G^2\rangle$ being a lower bound
  translational order, only data for $T>0.8$ are shown.}
\label{fss_trans}
\end{center}
\vskip 0.4 in
\end{figure}

In Figure~\ref{fss_trans} we show the approximate validity of
finite-size  scaling  by plotting the
  dimensionless quantity $L^{\eta} \langle \Psi_{G}^2 \rangle$ versus
  $\ln(L/\xi)$, i.e.,   the logarithm of the ratio of the
  finite-system-size to the measured correlation length, for two different size
  systems using the lower bound of $\eta=0.25$
  according to the KTHNY theory, namely $1/4 < \eta < 1/3$. The
  best collapse is obtained for the parameters $A=0.02192$, $B=4.89$ 
and $T_m=0.61$ shown as the top curve in Figure~\ref{fss_trans}.
(Note that we have used the constraint $T_{m}>0.6$ as indicated by the
behavior of the critical exponent $\eta$).
Using the values of the parameters obtained for this ``best'' collapse we obtain the
curve for $\xi(T)$ shown as a green line in Fig.~\ref{data_xi}.
The collapse obtained by using the parameters obtained by
the best fit to the correlation length (corresponding to the red curve
in Fig.~\ref{data_xi})  is shown as the graph at the bottom. We have
also included two more fits of both types of data, obtained by using parameter values between the 
above two extremes. We can observe that while we 
do not obtain the best fit of both sets of data (i.e., collapse of
$L^{\eta} \langle \Psi_{G}^2 \rangle$ versus
  $\ln(L/\xi)$ (Fig.~\ref{fss_trans}), and the temperature dependence of 
$\xi(T)$ (Fig.~\ref{data_xi}) for the same values of these parameters we
see that the values of $T_m$ and $b$ are close, only the prefactor $A$ 
cannot be accurately determined. We feel that the overall quality of fit
is reasonable given the fact that we had the difficulty discussed
  above in determining the correlation length associated with
  translational order.  

Several experimental  investigations \cite{Murray,Tang} have used the  decay of
the envelope of  $g(r)$ to extract $\xi$. The  resulting values of $\xi$
appear not to diverge across  the melting transition, so perhaps there
is some  shortfall in  using $g(r)$ to  get $\xi$ at  low temperature.
For instance,  Murray and Van Winkle  observe a finite  peak in $\xi$,
while for $\xi_6$ a divergence is seen to occur \cite{Murray}.

\begin{figure}
\vskip 0.3 in
\includegraphics[width=\figwidth]{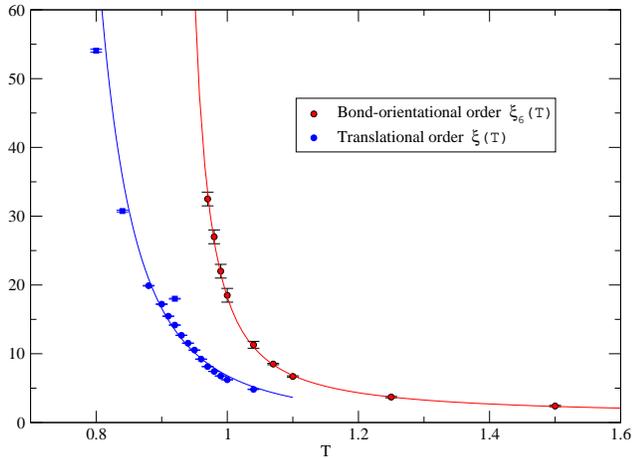}
\caption{Correlation lengths of the translational order parameter
are compared to $\xi_6(T)$.}
\label{compare}
\vskip 0.3 in
\end{figure}

Regardless of these differences, if we plot the results for $\xi(T)$
on the same plot with the results for $\xi_6(T)$ as shown in 
Figure~\ref{compare}, we see clearly that these two correlation length
are very different and the differences between these various fitting
forms for $\xi$ are not significant on this scale.

\section{Binder ratios}
\label{binder}

A central concept in finite-size scaling theory is that any 
dimensionless quantity should be a function of dimensionless ratios of 
the finite-size 
length ($L$) of the system to the correlation length $\xi(T)$ which 
emerges naturally
and it diverges near the critical point~\cite{Privman90}. 
Therefore, close enough to the critical point a dimensionless quantity
becomes a scaling function $f(L/\xi)$. At precisely the critical point, 
where the correlation length diverges, all dimensionless quantities 
are expected to be independent of the system size.

A straightforward way to construct a dimensionless variable is to take 
the ratio of cumulants. A simple non-trivial ratio is the so-called 
Binder ratio~\cite{Binder81} of the fourth and second cumulants,
\begin{equation}
U(x)=1-\frac{\langle (x-\langle x\rangle )^{4}\rangle }{3\langle (x-\langle x\rangle )^{2}\rangle ^{2}}.
\end{equation}
As mentioned above, this( dimensionless) variable is expected to be 
system-size independent at a critical point. Hence, if the values of 
$U(x)$ for several system sizes are plotted across a continuous phase
transition, they should cross at the critical point. This is the 
standard way of estimating for example the critical temperature of 
a thermal phase transition using the method of Binder ratios.

In the case of melting in two dimensions, however, we have seen that there are 
two correlation lengths: one for translational order, and another for 
bond orientational order. Clearly, if we approach very close to either
$T_i$ or $T_i$ only one of these two correlation lengths dominates. For example
if we approach $T_m$ sufficiently close from above, $\xi$ becomes very
large and $\xi_6$ is infinite. Thus, there is only one finite correlation 
length.  When we approach $T_i$ from above, both $\xi_6$ and $\xi$ are
finite, but if we are sufficiently close to $T_i$, $\xi_{6} \gg \xi$ and so
we can neglect the influence of $\xi$. 
In practice, however, because $\xi_6$ grows very rapidly   as the temperature
$T_i$ is approached and we can only study finite-size size systems, the
size of $\xi$ is not necessarily negligible as compared to the size of $\xi_6$.
This implies that  the scaling function becomes 
 $f(L/\xi, L/\xi_{6})$. As we have shown in the 
previous section, $\xi$ is still finite when $\xi_{6}$ diverges at the 
upper critical temperature, $T_{i}$. Thus, the Binder ratio would 
only be expected to have a crossing at $T_{i}$ if $\xi\ll L$, 
which is not the case for the system sizes we have considered 
($\xi(T_{i})\approx20$, half the length of the smallest system size). 
However, depending on the exact form of the scaling function, there 
may still be a crossing in the vicinity of $T_{i}$.

\begin{figure}
\vskip 0.3 in
\includegraphics[width=\figwidth]{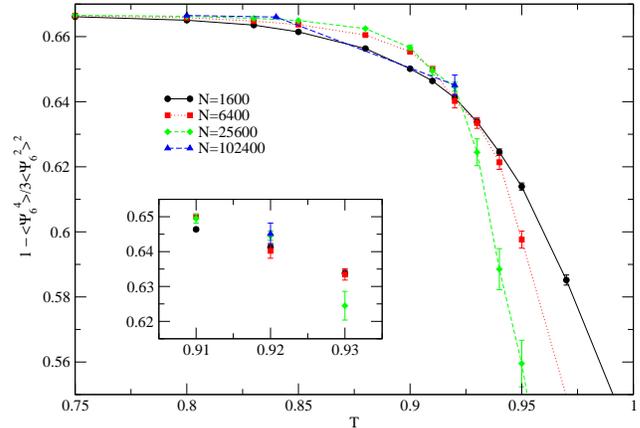}
\caption{Simplified~\cite{Keola10} Binder ratio of the bond orientational 
order parameter (lines are guides for the eye). 
The inset shows $U(\Psi_{6})$ near the crossing temperature, $T_{c}$.}
\label{abind}
\vskip 0.3 in
\end{figure}

Looking at the Binder ratio in Figure~\ref{abind}, we can see that 
there is an apparent crossing of $U(\Psi_{6})$ at  $T_{c}=0.92\pm0.01$. 
Although our statistical uncertainty is too great to identify the system-size 
dependence of the Binder ratio crossing (see inset), 
the finite-size scaling theory outlined 
above indicates that $T_{c}$ should approach $T_{i}$ when $\xi\ll L$. 
Thus, while we could use the value of $T_{c}$ as an estimate of $T_{i}$, 
the method obtained earlier for critical exponents is expected to yield 
more accurate results to the system sizes considered here. 

While we have also calculated $U(\Psi_{G})$, the shortfalls of our 
estimator for translational order in the temperature region
$0.6 \le T \le 0.8$ lead to an inconclusive analysis of the Binder crossing. 

Lastly, let us point out that the finite-size scaling theory discussed 
in this section should be applicable to any dimensionless parameter. 
In Sec.~\ref{correlations} we demonstrated the scaling collapse of 
the quantity $L^{\eta_{6}} \langle \Psi_{6}^2 \rangle$ when plotted 
as a function of $L/\xi_{6}$. In light of the analysis above, it is clear 
that we have neglected the $\xi$ dependence of this dimensionless 
quantity. In Figure~\ref{data_xi} we can see that for $T>0.95$ $\xi$ 
is more or less constant ($\xi\approx 8$). But as $T_{i}$ is approached, 
$\xi$ increases more rapidly, such that $\xi(T_{i})\approx 20$. 
This could explain the scatter seen in the scaling collapse of  
 $L^{\eta_{6}} \langle \Psi_{6}^2 \rangle$  shown in Figure~\ref{fss_bond}.

\section{Conclusions}
\label{conclusions}

We have  shown that several key  predictions from the  KTHNY theory of
two-stage  continuous melting  are  seen in  the  classical system  of
Lennard-Jones (LJ)  particles in two  dimensions. 

First, using  Delaunay triangulation we can define disclinations and
dislocations and this allows us to investigate the role of
defects in the 2D melting process. We can clearly observe at low
temperature that disclinations of 5-fold coordinates atoms and
disclinations of 7-fold coordinated atoms are bound into dislocations
which themselves are bound into dislocation pairs. Near $T_m$ we
begin to see unbound dislocations and at a higher temperature we begin
to observe unbinding of disclinations. The derivative with respect to
temperature of the total defect fraction exhibits a broad peak near $T \sim
0.9 $ very similar to the specific heat peak. Near this
temperature we find that the short-range peak (main peak) of the pair distribution
function of the 5-fold coordinated atoms and that of the 7-fold
coordinated atoms greatly diminishes. The pair distribution function
of 5-fold-7-fold coordinated particles also decreases greatly at roughly the
same temperature.

We calculated the distribution of the order parameters $\Psi_G$ and 
$\Psi_6$ on the complex plane. Below $T_m$, we see the characteristic ``Mexican
hat''-like circularly symmetric distribution for $\Psi_G$, i.e., while
the magnitude of $\Psi_G$ is finite below $T_m$, its phase fluctuates
causing the system to lose its translational order. The orientational
order parameter, $\Psi_6$, however, remains frozen to a particular
direction below $T_m$ because the system is large enough to allow, for
all practical purposes, for such a spontaneous symmetry breaking. In the
temperature range $T_m<T<T_i$, the ``Mexican-hat''-like distribution
of $\Psi_G$ collapses to a distribution around zero, while the
distribution of $\Psi_6$ becomes ``Mexican-hat''-like.
This could serve as a textbook description  of the hexatic order. For
$T>T_i$ the distribution of both $\Psi_G$ and $\Psi_6$ are centered
around zero value.

 We also calculated the temperature dependence of the second
 moment of the above two order parameters for various size systems
and from the size-dependence of the results  we have extracted the
anomalous dimensions, i.e., the critical exponents $\eta$ and $\eta_6$. 
 
Furthermore, we calculated the correlation functions $C_G(\vec r)$ and
$C_6(\vec r)$ of the order parameter $\Psi_G$ and $\Psi_6$
respectively. We find that both are controlled by two characteristic
correlation lengths, one is $\xi(T)$, which  characterizes the decay of the
correlation of the atomic positions and the other is $\xi_6(T)$, which
provides  the decay of the bond-orientation correlations.  We
demonstrate that we can  accurately extract both $\xi(T)$ and
$\xi_6(T)$.

We find that the two correlation lengths $\xi_6(T)$ and $\xi(T)$ have
very different temperature dependence, each diverging as we lower the
temperature at two different characteristic critical temperatures
$T_i$ and $T_m$ respectively, obtained by fitting the calculated
correlation length to the forms suggested by KTHNY theory. Furthermore,
  using the calculated correlation length and critical exponent
$\eta_6$, we find that the dimensionless quantity 
$L^{\eta_6} \langle \Psi_6^2 \rangle$
as a function of the dimensionless ratio $\ln(L/\xi_6)$ for all
size-lattice considered here collapse onto the same scaling
function. A similar conclusion is also reached for the 
finite-size scaling of the corresponding quantities related to
the translational order, i.e., 
 $L^{\eta} \langle \Psi_{\vec G}^2 \rangle$ versus
$\ln(L/\xi)$. This provides additional support for the KTHNY theory.

\end{document}